\documentclass[12pt,preprint]{aastex}

\shorttitle{SN\,2006gy: was it really extra-ordinary?}

\shortauthors{Agnoletto et al.}

\begin{document}


\title{SN\,2006gy: was it really extra-ordinary?}


\author{I. Agnoletto\altaffilmark{1}}
\affil{Dipartimento di Astronomia, \\ Universit\`{a} degli studi di
Padova, Vicolo dell'Osservatorio 3, I-35122, Padova, Italy}
\email{irene.agnoletto@oapd.inaf.it}

\author{S. Benetti, E. Cappellaro, L. Zampieri}
\affil{INAF-OAPD, Vicolo dell'Osservatorio 5, I-35122, Padova,
Italy}

\author{M. Turatto}
\affil{INAF-OACT, Via S.Sofia 78, 95123, Catania, Italy }

\author{P. Mazzali\altaffilmark{1}}
\affil{Max-Planck Institut f\"{u}r Astrophysik,
Karl-Schwarzschildstrasse 1, D-85741 Garching bei M\"{u}nchen,
Germany}

\author{A. Pastorello}
\affil{Astrophysics Research Centre, Physics Building, Queen's
University, BT7 1NN, Belfast, UK}

\author{M. Della Valle\altaffilmark{5,6}}
\affil{European Southern Observatory, Karl-Schwarschild-Strasse 2
D-85748 Garching bei M\"{u}nchen, Germany}

\author{F. Bufano, A. Harutyunyan, H. Navasardyan}
\affil{INAF-OAPD, Vicolo dell'Osservatorio 5, I-35122, Padova,
Italy}

\author{N. Elias-Rosa, S. Taubenberger}
\affil{Max-Planck Institut f\"{u}r Astrophysik,
Karl-Schwarzschildstrasse 1, D-85741 Garching bei M\"{u}nchen,
Germany}

\author{S. Spiro}
\affil{Dipartimento di Fisica, Universit\`{a} di Roma Tor Vergata,
Via della ricerca scientifica 1, I-00133, Roma, Italy}

\and

\author{S. Valenti}
\affil{Astrophysics Research Centre, Physics Building, Queen's
University, BT7 1NN, Belfast, UK}

\altaffiltext{1}{INAF-OAPD, Vicolo dell'Osservatorio 5, I-35122,
Padova, Italy}

\altaffiltext{5}{INAF - Osservatorio Astronomico di Capodimonte,
Salita Moiariello, 16 80131 Napoli, Italy}

\altaffiltext{6}{Center for Relativistic Astrophysics Network,
Piazza della Repubblica 10, I-65122 Pescara, Italy}

\begin{abstract}

We present the photometric and spectroscopic study of the very
luminous Type IIn SN 2006gy for a time period spanning more than one
year. The evolution of multiband light curves, the pseudo-bolometric
(BVRI) light curve and an extended spectral sequence are used to
derive constraints on the origin and evolution of the SN.

A broad, bright ($\rm{M_R}\sim-21.7$) peak characterizes all
monochromatic light curves. Afterwards, a rapid luminosity fading
($\gamma_{R}\sim3.2$ $\rm{mag\,(100\,d)^{-1}}$) is followed by a
phase of slow luminosity decline ($\gamma_R\sim0.4$
$\rm{mag\,(100\,d)^{-1}}$) between day $\sim$170 and $\sim$237. At
late phases ($>237$ days), because of the large luminosity drop
($>3$ mag), only upper visibility limits are obtained in the B, R
and I bands. In the near-infrared, two K-band  detections on days
411 and 510 open new issues about dust formation or IR echoes
scenarios.

At all epochs the spectra are characterized by the absence of broad
P-Cygni profiles and a multicomponent $\rm{H\alpha}$ profile, which
are the typical signatures of type IIn SNe. $\rm{H\alpha}$
velocities of $\rm{FWHM\approx 3200}\,\rm{km\,s^{-1}}$ and
$\rm{FHWM\approx 9000}\,\rm{km\,s^{-1}}$ are measured around maximum
phase for the intermediate and high velocity components,
respectively, and they evolve slowly with time.

After maximum, spectroscopic and photometric similarities are found
between SN 2006gy and bright, interaction-dominated SNe (e.g.
SN\,1997cy, SN\,1999E and SN\,2002ic). This suggests that ejecta-CSM
interaction plays a key role in SN\,2006gy about 6 to 8 months after
maximum, sustaining the late-time-light curve. Alternatively, the
late luminosity may be related to the radioactive decay of
$\rm{\sim3M_{\odot}}$ of $\rm{^{56}Ni}$.

Models of the light curve in the first 170 days suggest that the
progenitor was a compact star ($\rm{R\sim6-8\cdot10^{12}cm}$,
$\rm{M_{ej}\sim5-14M_{\odot}}$), and that the SN ejecta collided
with massive ($\rm{6-10M_{\odot}}$), opaque clumps of previously
ejected material. These clumps do not completely obscure the SN
photosphere, so that at its peak the luminosity is due both to the
decay of $\rm{^{56}Ni}$ and to interaction with CSM. A supermassive
star is not required to explain the observational data, nor is an
extra-ordinarily large explosion energy.

\end{abstract}



\section{Introduction}

Textbook stellar evolution theory explains that Type II SNe are
produced by the core collapse of H-rich stars with masses $\gtrsim 8
{\rm M}_{\odot}$ \citep{bra90,arn96}. Confirmation of this scenario
comes from the possible identification of the progenitors of a few
SNe\,II, including SN\,1987A \citep{arn89} and several more recent
events \cite[e.g.][]{smart04,vandik}.

On the other hand, a fully consistent picture of a SN progenitor
evolution and explosion is still missing. Parameters like progenitor
radius, ejecta mass, explosion energy, asymmetries and radioactive
elements yield, all contribute to determining the SN display.

One of the most uncertain ingredients is the progenitor mass-loss history.
Although some constraints on the progenitor mass loss can be derived from models
of the SN light curves and spectra, direct measurements are only possible when
the circumstellar material (CSM) becomes visible.

Denser CSMs and hence higher mass loss rates
($\rm{\sim10^{-4}\,M_{\odot}\,yr^{-1}}$, \citealt{sal02}) are
required to explain the sudden halt in the late-time luminosity
decline of several type II Linear SNe (e.g., SN\,1979C,
\citealt{bra81}; SN\,1980K, \citealt{mont98}; SN\,1986E,
\citealt{ec95}).

Mass loss plays a key role in type IIn SNe \citep{schl90}. The
spectra of this class of SNe are characterized by emission lines
with multiple components, which may range from very high ($\sim20000
\,{\rm km\,s}^{-1}$) to low velocities (a few hundred ${\rm
km\,s}^{-1}$) with no associated (broad) P--Cygni absorptions. Such
events may be very energetic (e.g., \citealt{arx99}) and are
characterized by a slow luminosity evolution beginning soon after
discovery. It is generally believed that the shock produced when the
high velocity ejecta impact on a relatively dense CSM causes the
conversion of part of the ejecta kinetic energy into radiation.
Depending on the CSM density distribution, the strong CSM-ejecta
interaction may last for months (e.g., SN\,1994W, \citealt{chu04})
to years (e.g., SN\,1988Z, \citealt{mt93}; SN\,1995N,
\citealt{fran02} and \citealt{lz05}; SN\,1995G, \citealt{apas02}).

The recent peculiar SN IIn 2006gy event attracted much interest.
Discovered on 2006, September 18th in NGC 1260 \citep{qui06},
SN\,2006gy was initially classified as a SN\,II \citep{avik06} and
shortly thereafter as a SN\,IIn \citep{fo07}. Although its
spectroscopic features were not unprecedented, the photometric
behavior was. A very bright luminosity peak ($\rm{M_{R}\sim -22}$,
\citealt{sm07}, hereafter S07) was in fact reached only $\sim 70 $
days after explosion (estimated by backward extrapolation of the
light curve, cr. \S2). The amount of energy radiated in visible
light during the first 200 days ($\rm{>10^{51}\,erg\,s^{-1}}$, S07)
was larger than in any previously observed SN, either core-collapse
or thermonuclear.

On the other hand, in contrast with other bright SNe\,IIn
(\citealt{lz05}, \citealt{cha05}), only a weak and soft X--ray
emission was detected by \textit{Swift} and \textit{Chandra} near
the epoch of optical maximum (S07). The absorption-corrected,
absolute X-ray luminosity in the band $0.65-2$\,keV was $1.65 \times
10^{39}\,{\rm erg\, s}^{-1}$ (S07). It was argued (\citealt{of07},
S07) that if direct radiation from the ejecta-CSM interaction was
the cause of the extraordinary SN luminosity, the X-ray flux should
have been several orders of magnitude larger. A possible explanation
is that the interaction region might have been hidden because of a
very large optical depth. Alternatively, the source for the optical
luminosity should be searched elsewhere.

Actually, the first model proposed by S07 was the explosion of a
very massive star ($\rm{>100 M_\odot}$) via
\textit{pair--instability} phenomena. Such a violent explosion would
cause the complete disintegration of the core and the ejection of a
huge amount of $\rm{^{56}Ni}$, possibly up to 22M$_\odot$. In such a
scenario the energy input from the radioactive decay chain
$\rm{^{56}Ni\rightarrow^{56}Co\rightarrow^{56}Fe}$ accounts for the
observed luminosity.

However, late luminosity measurements presented by \citet{sm08}
(hereafter S08) and in this work set a much estimate to the possible
$\rm{^{56}Ni}$ mass. Other models call for the conversion of some of
the ejecta kinetic energy into radiation, via a collision with a
massive ($\rm{\sim 10M_\odot}$) and highly opaque ($\tau \sim$300)
circumstellar shell (\citealt{smcray07} and S08). This scenario
provides no direct information on the real nature of the explosion
which, in principle, could even be a thermonuclear runaway
\citep{of07}. However, this is difficult to reconcile with the
presence of the massive circumstellar shell. Thus, the core collapse
of a massive star is still the most appealing scenario.
Alternatively, we may have witnessed the collision between high
velocity shells originated in subsequent outbursts of a very massive
star undergoing structural instabilities caused by pair production
(\textit{pulsational pair-instability}, \citealt{woos07}). In any
case, the presence of massive shells ($\rm{\sim 10 M_\odot}$) at a
large radius (i.e., a few $10^{15}$ cm) is required to explain the
observed high luminosity. Such a large shell mass, coupled with a
high density, may result in long photon diffusion times
(\citealt{smcray07}), which could explain the broad peak of the
light curve of SN\,2006gy.

These scenarios can explain the light curve evolution during the
first 5 months. However, the presence of a diffusion process implies
fairly rapid fading of the luminosity after maximum (i.e., a time of
the order of the photon diffusion time-scale).  Observations instead
indicate that the luminosity does not decline rapidly after maximum.
On the contrary, the light curve decline slows down between day 170
and day 237 (see \S 3). This requires an additional energy source.
Again, radioactive material has been proposed ($\rm{\sim8
M_{\odot}}$ of $\rm{^{56}Ni}$ according to \citealt{smcray07}), but
new doubts have arisen after S08 reported a late-time drop of the
optical luminosity and two infrared SN detections, which may support
the formation of dust in the ejecta or the presence of IR echoes.

In this work we present new data, which include observations at all relevant
phases, from discovery to more than 1 year later. Multiband light curves and an
extended spectral sequence are shown and discussed. By comparison with other SNe
and by means of modeling we try to verify to which extent this SN is really
extra-ordinary and to provide new constraints for the progenitor and the
explosion.

\section{Observations}

Optical (BVRI) and near--infrared (JHK') images of SN\,2006gy were
acquired at TNG, NOT (La Palma, Spain) and the Copernico 1.82m
Telescope on Mt.\,Ekar (Asiago, Italy) over a period spanning more
than 500 days from discovery. Optical spectroscopy was also
performed, up to $\sim$389 days (see Table 1 for a complete log of
the observations).

Since the SN is located very close to the nucleus of the host galaxy,  template
subtraction was required for photometry. We used archival B, R and I band images
of NGC1260 acquired at the Jakobus Kapteyn Telescope\footnote{web archive
http://archive.ast.cam.ac.uk/ingarch/ingarchold.html} (La Palma, Spain), and a V
band image taken at the Schmidt Telescope at Kiso Observatory\footnote{web
archive http://www.ioa.s.u-tokyo.ac.jp/kisohp/} (Japan). Additional information
about the template images is given in Table 2.

All images were de-biased and flat--field-corrected. A local
sequence of stars in the SN field was calibrated using observations
of standard stars obtained during photometric nights. Template
subtraction was performed using ISIS \citep{alard98}, and the SN
magnitudes were measured on the subtracted image with a Point Spread
Function-fitting technique, using custom-made DAOPHOT-based routines
that were adapted specifically for supernova photometry.

For spectroscopy, all scientific exposures were acquired at low
airmass and positioning the slit along the parallactic angle.
Wavelength calibration was accomplished with arc-lamp exposures and
checked against the night-sky lines. The flux was calibrated using
instrumental sensitivity functions obtained from observations of
spectrophotometric standard stars. These were used also to remove
telluric absorptions from the spectra. In order to improve the
signal-to-noise ratio, separate spectra taken during the same nights
were combined. Finally, flux calibration was checked against
photometry. If necessary, a constant multiplicative factor was
applied to correct for flux losses caused by slit miscentering or
non--photometric sky conditions.

The spectrum acquired on 2006 December 19th with the Ekar 1.82m
telescope required further adjustments because of the poor seeing
conditions and the residual contamination from the galaxy
background. The latter was removed using the spectrum at 389 days,
where the SN is not detected, as a background template.

Throughout this paper, for the sake of simplicity, phase refers to
the same \textit{reference epoch} as in S07, JD=2453967 (2006 August
19.5 UT). This was derived from a backward extrapolation of the
rising branch of the light curve. S07 refer to this date as the
explosion epoch, but this term may be misleading in view of some of
the proposed scenarios\footnote{In the shocked-shell diffusion model
\citep{smcray07} the detection of the SN emission occurs a few weeks
after the real explosion. In the scenario suggested by
\citet{woos07} there is not even a SN explosion, but only an
outburst release of matter.}.

\section{Photometry}

To compute the absolute magnitudes of SN\,2006gy some assumptions
regarding the host galaxy distance and SN extinction are needed.

Lacking other indicators, the distance to NGC 1260 was estimated
from the Hubble law, $\rm{d=v_{rec}/H_0}$, where $\rm{v_{rec}=5822km
s^{-1}}$ is the host galaxy recession velocity corrected for Virgo
cluster infall (from
\emph{HyperLEDA}\footnote{http://leda.univ-lyon1.fr/}) and
$\rm{H_{0}}=72\pm8 ~{\rm km\, s}^{-1}\, {\rm Mpc}^{-1}$
\citep{fr01}. These values imply a distance modulus $\mu=34.53$,
equivalent to a distance of 80.86 Mpc.

We adopted a Galactic extinction towards NGC~1260 of
$\rm{E(B-V)_{gal}=0.16}$ ($\rm{A_{B, gal}=0.69}$, \citealt{schl98}).
An estimate of the extinction in the host galaxy was obtained
comparing the spectra of SN\,2006gy to those of SN\,II 2007bw,
another peculiar and bright SN\,IIn,  photometrically similar to
SN\,2006gy. This yields $\rm{E(B-V)_{host} \simeq 0.4}$, assuming
for SN\,2007bw little or no extinction. It is interesting to note
that on the 42 days spectrum the EWs of NaID due to the galactic and
interstellar absorption are 2.2$\rm{{\AA}}$ and 5.5$\rm{{\AA}}$,
respectively. Assuming for NGC 1260 a gas-to-dust ratio along the
line of sight as in our Galaxy and adopting the extinction of
\citet{schl98}, the derived internal absorption is fully in
agreement with that derived by comparison with SN\,2007bw.


Therefore the total color excess is $\rm{E_{tot}(B-V) \simeq0.56}$.
This is comparable to the estimate of S07, i.e. $\rm{E_{tot}(B-V)
\simeq0.48}$, which was also obtained by comparison with SNe\,IIn,
and slightly smaller than the value adopted by \citet{of07},
$\rm{E_{tot}(B-V) \simeq0.7}$.

The BVRI absolute light curves of SN\,2006gy are plotted in Figure
1. In all bands the light curve exhibits a slow increase to maximum,
which is reached $\sim 70$ days after the reference epoch. The peak
magnitudes are $\rm{B \sim-21}$, $\rm{V \sim-21.4}$, $\rm{R
\sim-21.7}$ and $\rm{I \sim-21.5}$. Such an extended, plateau-like
peak was noted for a  type IIn SN only in the case of SN 2005kd
\citep{tsv08}.

Between day $\sim100$ and day $\sim170$, the light curve declined
relatively rapidly ($\gamma_{B}\sim3.0$$\rm{mag\,(100\,d)^{-1}}$,
$\gamma_{V}\sim3.1$, $\gamma_{R}\sim3.2$, $\gamma_{I}\sim2.8$).
Then, from day $\sim170$ onwards the light curve evolution suddenly
flattened: in the following $\sim70$ days the decline was only
$\gamma_{R}\sim0.4$$\rm{mag\,(100\,d)^{-1}}$. When the SN could be
observed again after solar occultation, its luminosity was below the
detection limit in the optical bands. A limit was obtained placing
artificial stars of different magnitudes at the SN position. Despite
the long exposure times, only relatively bright upper limits were
derived because the SN is very close to the nucleus of the galaxy
(cf. S07, Figure 1). The derived apparent magnitude limit on day 389
is $\gtrsim 20.3$ in R. Optical upper limits were obtained also at
423 days, when we derive B $\gtrsim 21.0$, R $\gtrsim 21.5$ and I
$\gtrsim 19.75$. These measurements imply a new steepening in the
luminosity decline after day 237.

Guided by the evolution of other SNe (e.g., SN\,1998S,
\citealt{poz04} or SN\,2006jc,
\citealt{tom07,sm08c,sep08,dcarlo08}), we considered the possibility
that at late epochs a significant fraction of the bolometric
luminosity could be emitted at IR wavelengths. To test whether  this
was the case, late observations of SN\,2006gy were obtained with
NICS at the TNG, on 2007, October 5 (JHK' bands) and on 2008,
January 12 (K' band only).  We could not apply the template
subtraction technique in the near-infrared because the available
pre-discovery images of the host galaxy retrieved from the 2MASS
archive are not deep enough to be compared to the TNG images.
Therefore, we had to rely on the PSF fitting technique which, given
the SN position, has a large uncertainty. Photometric calibration
was performed adopting field star magnitudes as listed in the 2MASS
Point Source
Catalogue\footnote{http://tdc-www.harvard.edu/software/catalogs/tmpsc.html}.

The SN was not detected in the J and H bands, for which we could
only estimate upper limits, J $\gtrsim$ 17.0 and H $\gtrsim$ 16.5.
Instead, a point source was detected in the K' band at the SN
position (Figure 2). The SN was measured at K $\sim16.0\pm 0.5$ on
day 411 and K $\sim16.3\pm0.5$ on day 510. These values are $\sim
1$\,mag fainter than those measured by S08 at similar epochs
(K=$15.1\pm0.1$ and K=$15.4\pm0.1$ on day 405 and 468,
respectively). Even allowing for the large error bars, the two sets
of measurements do not agree, probably because of a different
calibration.

\subsection{Infrared emission and bolometric luminosity}

Given the K'-band detection of the SN at 411 and 510 days, it cannot
be excluded that at late phases a considerable amount of flux is
emitted in the near-infrared.

S08 suggested two possible sources of the late K'-band luminosity.
The K'-band emission could be associated with an IR light echo from
circumstellar dust, for which the input energy is the light emitted
by the SN near maximum. In this case the IR flux should not be
considered when computing the late-time bolometric light curve.
Alternatively, the IR flux may originate from circumstellar dust
heated by an instantaneous energy supply (radioactive decay or
on-going CSM-ejecta interaction), as was suggested by \citet{poz04}
to explain the late phase photometric data of SN\,1998S.

In order to get some constraints on the total emission from dust at
$\sim 411$-423 days we assumed a black body energy distribution
multiplied to a factor $\rm{1/\lambda}$, as an approximation of what
reported in \citet{spi98}, and normalized it to the observed K-band
flux. Given that we have no constraints on the dust temperature, we
adopted three values including T=1200K, the dust temperature in the
ejecta of SN 1998S derived by \citet{poz04}. For each value we
plotted the spectral energy distribution (SED) of the associated
emission (Figure 3) and integrated over the entire wavelength range
from $\rm{\lambda_K=2.16\,{\mu}m}$ to $\lambda=\infty$.

It is interesting to note that the K-band luminosity of SN\,1998S
measured at similar epochs (K=13.8 at $\sim$464 days) would differ
from that of SN\,2006gy by a factor 1-5 (adopting for SN\,2006gy
K=16.3 from this work or K=15.4 from S08) if scaled at the same
distance. Therefore, given that the two fluxes are of the same
order, it is plausible that any mechanism explaining the IR emission
of SN\,1998S can work also for SN\,2006gy.

From our multiwavelength photometry we can derive the
pseudo-bolometric luminosity evolution of SN\,2006gy integrating the
flux in the optical bands (BVRI). The pseudo-bolometric light curve
is shown in Figure 4, compared to those of the type II SNe\,1987A,
1995G, 1997cy, 1999E and 2005gj. The pseudo-bolometric luminosities
which include the dust contribution in the near IR are represented
with plus symbols. It is remarkable that, at about 6 to 8 months,
the luminosity and decay rate of SN\,2006gy become comparable to
those of other events, in particular to SNe\,1997cy and SN\,1999E.
We will come back to this issue in $\S4$ and $\S5.2$.

\section{Spectroscopy}

The spectral evolution of SN\,2006gy from day 37 to day 389 is
presented in Figure 5. The spectra have been de-redshifted and
corrected for extinction.

The early spectra (days 37 and 42) show the typical features of SNe\,IIn, namely
a blue continuum and strong H$\alpha$ and H$\beta$ emission lines, without broad
P-Cygni absorptions.

Using \emph{GELATO} \citep{avik08}, the automatical spectra
comparison tool applied to the Asiago Supernova Archive (ASA), the
best match for the earliest available SN\,2006gy spectrum (phase 37)
is found with SN\,1995G ($\sim$36 days since discovery,
\citealt{apas02}), which is generally considered a typical SN\,IIn,
although there are differences in the strength of the lines
emission.

Three months after discovery, the spectrum of SN\,2006gy became
similar to those of other well-studied SNe\,II. In the high
resolution spectrum taken on day 96 by S07 (their Figure 4) narrow
absorption lines of Fe II (multiplets 42, 48, and 49 at
5000-5400\,\AA\/ and multiplet 74 in the region 6100-6500\,\AA) are
evident. A similar narrow line forest was identified in the spectra
of SN\,1999el \citep{dcarlo02}, SN\,1995G \citep{apas02} and
SN\,1994W \citep{soll98,chu04}. In all cases these lines are
associated with slowly expanding, unperturbed material surrounding
the star.

Despite the lower resolution and S/N ratio, our spectrum at phase 122 days is
broadly consistent with the features discussed by S08. At this phase the
H$\alpha$ flux has decreased by a factor of 2 with respect to the first
spectrum, while H$\beta$ emission almost disappeared.

At a phase of 174 days the near-IR Ca II triplet is strong in
emission. For this epoch a good spectral match is obtained with
SNe\,1997cy, 1999E (Figure 6), 2002ic and 2005gj.

The H$\alpha$ flux continues to decrease with time: on day 174 it is
$\sim$3 times fainter than on day 37 and on day 204 even 5 times.
Finally, the last spectrum (day 389) shows no evidence of the
typical lines of SNe\,II; at this epoch, the narrow H$\alpha$
emission should be attributed to the host galaxy. This is consistent
with S08 and with the upper limit in the optical luminosity that was
deduced from the photometry.

The emission peak of  $\rm{H{\alpha}}$ remains at the rest frame
wavelength at all phases, exhibiting a three--component profile
(Figure 7). For the intermediate $\rm{H\alpha}$ component we
measured a FWHM$\rm{\sim2100\,km\,s^{-1}}$ at a phase of 42 days,
and FWHM$\rm{\sim3200\,km\,s^{-1}}$ at 174 days. S07 pointed out an
asymmetry of the line at early times (also evident in our day 42
spectrum), likely caused by a blueshifted P-Cygni absorption, which
vanishes with time. For this reason we can admit that the true
unabsorbed profile H$\alpha$ remained roughly constant during the SN
evolution. A roughly constant FWHM$\sim9100\rm{km\,s^{-1}}$ is
measured for the broad component of $\rm{H\alpha}$.

The physical interpretation of the intermediate and broad components
is still a matter of debate. S07 and S08 assumed that the
intermediate component ($\rm{v\sim4000\,km\,s^{-1}}$) traces the
kinematics of the SN shock wave, while the broader one is related to
the SN ejecta  ($\rm{v\sim6000\,km\,s^{-1}}$, a value significantly
lower than what we obtain, $\rm{v\sim9100km\,s^{-1}}$). The
\textit{intermediate} velocity component was used to compute the
luminosity expected from CSM interaction. On the other hand,
according to \citet{chev94,chev01} and to \citet{lz05}, the
luminosity originating from the reverse shock during ejecta-CSM
interaction is proportional to the ejecta velocity, i.e. to the
width of the \textit{broadest} H$\alpha$ component. There is still
no consensus on this issue. Because of these ambiguities, one should
be careful before assigning physical velocities to various regions
from just line widths, especially for objects with peculiar
individual features as are SNe\,IIn.

\section{Discussion}

As discussed in $\S3$, the light curve of SN\,2006gy shows three
distinct phases: \textit{i)} a very broad, exceptionally high
luminosity peak (day 0 to $\sim$170), \textit{ii)} an intermediate
phase of slow decline (day $\sim$170 to $\sim$237) and \textit{iii)}
a late phase in which the optical luminosity drops below the
detection limit and IR emission dominates (day $>$389). As we will
show, the first phase requires a specific star+CSM configuration.
The other two phases have been observed in other SNe.

In the following, we rewind the movie of the event and use the late
observations to constrain the possible scenario. Starting from the
late phases, we discuss the role of dust and $\rm{^{56}Ni}$ in the
ejecta, stressing that a very large amount of $\rm{^{56}Ni}$ is not
required. We then consider the evolution of the SN at intermediate
phases and explain that, independently of the source that powered
the luminosity at peak, interaction dominates between days $\sim
170$ and $\sim 237$. Finally, we discuss the light curve models for
the first $\sim$5 months obtained with a semi-analytical code
\citep{lz03}. Based on these results, we propose a new evolutionary
scenario for SN\,2006gy.

\subsection{Nickel mass and dust emission}

The late light curve of most SNe is powered by the radioactive decay
of $\rm{^{56}Ni}$ into $\rm{^{56}Co}$ and $\rm{^{56}Fe}$ via
$\gamma$ and $\rm{e^{+}}$ deposition.

Thermonuclear SNe\,Ia eject a large $\rm{^{56}Ni}$ mass
($\rm{0.1M_{\odot}<M_{^{56}Ni}<1.1M_{\odot}}$,
\citealt{ec97,maz07}), but become rapidly transparent to the
$\gamma$-rays from the radioactive decay because of the small
ejected mass and the high expansion velocity. As a consequence, at
$t \sim 100$ days after explosion the luminosity declines at a rate
$\sim$1.5$\rm{mag\,(100\,d)^{-1}}$, higher than the $\rm{^{56}Co}$
decay input ($\sim0.98\rm{mag\,(100\,d)^{-1}}$). A similar behavior
was found for most type Ib/c SNe \citep{clo97}.

In the case of H-rich, core-collapse SNe the ejecta remain almost
opaque to $\gamma$-rays for more than a year, and the late-time
luminosity decline tracks the radioactive decay. In this case, if
the date of the explosion is known, the late-time luminosity
provides a direct estimate of the ejected $\rm{^{56}Ni}$ mass. This
spans a wide range of values ($\rm{0.005<M_{^{56}Ni}<0.3M_{\odot}}$;
L. Zampieri et al., in preparation), but is typically smaller than
in SNe\,Ia and Ib/c.

For SN\,2006gy, in the optical bands only upper limits to the
luminosity at very late phases (411 days) can be obtained. In the
near infrared, the K-band detection reported in S08 and discussed in
the previous section may be suggestive of the presence of
low-temperature dust emitting in the far IR. This makes a precise
estimate of the ejected $\rm{^{56}Ni}$ mass difficult. The
bolometric luminosity including the emission from dust (plus symbols
in Figure 4) imply ejected $\rm{^{56}Ni}$ masses up to
$\rm{\sim15M_{\odot}}$ for T=800K. However, given the uncertainty on
the nature of dust and its temperature, a more significant estimate
of  $\rm{M(^{56}Ni})$ can be obtained adopting the bolometric
luminosity at earlier epochs, i.e. at $\sim180$ days. At this phase
the relation $\rm{L=1.4\cdot 10^{43}M_{Ni}\cdot
exp(-t/113.6)\,erg\,s^{-1}}$ provides $\rm{M_{Ni}\sim3M_{\odot}}$
assuming complete $\gamma$-ray trapping (see also $\S5.3$). This
value is in disagreement with the value obtained by \citet{smcray07}
with the same relation ($\rm{M_{^{56}Ni}\sim8M_{\odot}}$). A
possible explanation may reside in a different estimate of the
bolometric luminosity, which is about a factor 3 higher in
\citet{smcray07}.

Of course, we expect the bolometric flux - and therefore the
$\rm{^{56}Ni}$ mass - to increase if the IR/longer wavelength
emission contribution is taken into account. On the other hand, the
luminosity decay at phase 170-137 days is much slower than what is
expected from $\rm{^{56}Co}$ decay. This suggests that an energy
source additional to radioactive decay of $\rm{^{56}Ni}$ has to be
present. Compared to those measured for other SNe, an amount of
$\rm{3M_{\odot}}$ of $\rm{^{56}Ni}$ may not appear unreasonably
large (see for example SN\,1999as, \citealt{deng01}).

\subsection{Evidence of strong, late-time ejecta-CSM interaction}

We mentioned in $\S$3.1 and $\S$4 that at 170-237 days SN\,2006gy
shares several properties with SNe\,1997cy, 1999E, 2005gj and
2002ic. Although some authors regard some of these SNe as
thermonuclear explosions (see \citealt{ham03} for SN\,2002ic and
\citealt{pr07} for SN\,2005gj, but see also \citealt{sb06} and S.
Benetti et al. 2008, in preparation, for an alternative scenario)
there is unanimous consensus on the fact that interaction dominates
their emission at late phases. Despite the brighter magnitude at
maximum, SN\,2006gy has luminosity and luminosity decline rate
comparable to the SNe mentioned above at 170-237 days (Figure 4),
which is when they also show similar spectra.

Therefore, it is natural to assume that at this phase ejecta-CSM
\textit{interaction plays a dominant role also in SN\,2006gy.} Although the low
X-ray flux at this phase (cf. $\S$1) might appear to be in contradiction with
the ejecta/CSM interaction scenario, this may not be a problem, because for
sufficiently high densities ($\rm{\rho\sim10^{8}g\,cm^{-3}}$) the X-rays that
are produced in the shock are immediately absorbed \citep{mt00}.


In the context of interaction, the luminosity L arising from the
shock is proportional to the progenitor mass loss rate
$\rm{\dot{M}}$, to the ejecta velocity $\rm{V_{ej}}$ and to the
unshocked CSM wind velocity $\rm{V_{CSM}}$, as follows: $\rm{L
\propto V^3_{ej} \dot{M} V^{-1}_{CSM}}$. Unfortunately, because of
the ambiguity in the interpretation of emission line profiles
($\S$4), we cannot precisely measure the velocities in the different
circumstellar regions and thus  derive a reliable estimate of the
CSM density from the observational data. However, the emission lines
in SNe\,1997cy and 1999E are generally broader than in SN\,2006gy
(i.e., their ejecta are probably faster), but their luminosity is
comparable. In view of the former relation, we expect that the shock
wave of SN\,2006gy encounters a higher CSM density at 170-237 days.

\subsection{A highly energetic supernova impinging on massive gaseous clumps}

The SN evolution during the first 170 days is explained reasonably
well by the scenario proposed by \citet{smcray07}. In the
shocked-shell diffusion model the supernova light is produced by
diffusion of thermal energy after the passage of the SN shock wave
through a shell of $\rm{10M_{\odot}}$ of material, ejected in the
decade preceding the explosion. The shell is supposed to be
initially optically thick and acts as a pseudo-photosphere, so that
the long duration of the peak and the weakness of X-rays emission
are explained in terms of a long diffusion time and a very large
optical depth. The interaction features typical of type IIn SNe are
supposed to arise in the observed spectra as soon as the blast wave
breaks out of the opaque shell into the surrounding, lower-density wind.

However, in \citet{smcray07} a number of items have not been
considered. A first inconsistency concerns the model assumptions.
According to the model of \citet{arn73,arn77} which is adopted in
\citet{smcray07}, in order to reproduce the observed luminosity rise
to maximum, the initial radius of the shocked shell has to be much
smaller than the radius at peak luminosity. However, in the model of
\citet{smcray07} the initial and final radius differ only by a
factor of 2. Thus the model of \citet{arn73} is not applicable: the
simple assumption of the existence of a single shell at a large
radius surrounding the exploding star can not explain the properties
of the light curve of SN\,2006gy, in particular the slow rise to
maximum.

Secondly, in the model of \citet{smcray07} the important role of
$\rm{^{56}Ni}$ is overlooked. No attempt has been made to estimate
the amount of $\rm{^{56}Ni}$ deposited by the SN and to determine
its effect on the light curve during the diffusive
phase\footnote{The estimate of $\rm{8M_{\odot}}$ of $\rm{^{56}Ni}$
reported in their paper derives from the extrapolation of the light
curve luminosity after day 170 (cfr. $\S$5.1)}.

The third problem concerns recombination, whose effects can not be
neglected as soon as the decreasing photospheric temperature reaches
the gas recombination temperature during the post-diffusive phase.

With these shortcomings in mind, we have developed an alternative,
comprehensive scenario that attempts to take all these aspects into
account. First of all, we divided the evolution of SN\,2006gy into
two distinct phases, before and after maximum luminosity. Each phase
was modeled independently.  The earlier phase (i.e., the rising
branch of the light curve) was modeled as the explosion of a
core-collapse SN originating from a compact progenitor. For the peak
phase we adopted a scenario similar to that of \citet{smcray07}, in
which the ejecta impact on very massive ($\rm{>6M_{\odot}}$, see
Table 5) clumps of previously ejected material and deposit their
kinetic energy. Because the density is very high, the energy of the
shock produced by the ejecta-clump interaction is completely
thermalized. A photosphere forms, so that the evolution of the
shocked clumps can be modeled as if it was another SN with very
large radius and little ejected $\rm{^{56}Ni}$.

A fundamental difference with respect to the model of
\citet{smcray07} is that, in our scenario, the true SN explosion is
not completely hidden by the circumstellar material which is
therefore not homogeneously distributed around the star. Rather, it
is fragmented into big clumps which may be distributed symmetrically
with respect to the centre of the star. This is motivated by the
assumption that the progenitor of SN\,2006gy may have undergone
mass-loss episodes similar to those observed in $\eta$ Carinae.  The
rise to maximum corresponds to the early emission of the SN ejecta
during the initial phase in which the radius is rapidly expanding,
similarly to the case of SN\,1987A \citep{woos88}. In our model the
peak luminosity is sustained by the \emph{combined contribution} of
the early SN explosion and the energy from the ejecta-clump
interaction. Unfortunately, no early spectra are available to verify
this claim. The first available spectrum (37 days) already shows
signs of interaction, mainly in the H$\alpha$ profile, probably
caused by flux arising directly from the interaction, being not
thermalized by the dense clumps. Therefore we can reasonably assume
that at this phase the ejecta-clump collision had already started.
Another assumption of our model is that the impact is instantaneous,
i.e. that all material is reached by the ejecta at the same radial
distance from the star.

Our semi-analytical code (see \citealt{lz03} for more details) was
used to estimate the parameters of the ejected envelope from a
simultaneous comparison of the observed and computed light curve,
photospheric gas velocity and continuum temperature. The radius of
the star at the explosion, the mass and velocity of the ejecta and
the explosion energy are fitting parameters, whereas the ejected
$\rm{^{56}Ni}$ mass is an input fixed parameter, which is based on
the late-time light curve. The fitting parameters are estimated by
means of a $\chi^{2}$ minimization procedure for both evolutionary
phases (i.e. the SN explosion and the ejecta-clumps impact).

The parameters of the models for each phase are listed in Table 5
and 6. Models of the earlier phase (\textit{e1}, \textit{e2},
\textit{e3} and \textit{e4}) refer to different values of the input
parameters $\rm{M_{Ni}}$ and $\rm{T_{rec}}$, while models of the
later phase (\emph{c1} and \emph{c2}) refer to different $\chi^{2}$
minima.

Critical parameters for the earlier phase are the initial radius and
the mass of $\rm{^{56}Ni}$. As discussed before, the large increase
in luminosity in the pre-maximum phase calls for small initial radii
($<10^{13}$cm), which are not compatible with RSG stars but are
consistent with BSG or Wolf Rayet stars. The amount of
$\rm{^{56}Ni}$ determines the peak luminosity. The adopted upper
limit is $\rm{M_{Ni}\sim 2M_{\odot}}$, considering that $\rm{\sim
3M_{\odot}}$ were estimated at $\sim 180$ days neglecting the
contribution of interaction, which instead \emph{is} already active
at that phase, as explained above (according to the model
 the $\gamma$-rays trapping is always more than $80\%$ effective at
a phase 170-237 days, given the large ejecta masses).  A minimum
$\rm{^{56}Ni}$ yield of 0.75 $\rm{M_{\odot}}$ was required to fit
the early rise of the light curve, assuming no contribution from
interaction (i.e., the CSM is supposed to be rarified in the
vicinity of the exploding star).

Table 5 lists the best-fit-parameters for the initial radius of the
star, for the ejecta mass and for the velocity and SN explosion
energy, given the adopted $\rm{^{56}Ni}$ masses and recombination
temperatures. It should be noted that the radius estimated by the
code is actually an upper limit: the initial part of the light curve
is not very sensitive as long as it remains below the reported
values. The SN parameters are not exceptionally high for a
core-collapse SN. For example, the explosion energy is only
$\sim3-4.5$ times larger than that of SN\,1987A. The explosion
energy increases with increasing $\rm{^{56}Ni}$ mass, as one may
naively expect from the fact that larger amounts of $\rm{^{56}Ni}$
may be synthetized in more energetics events. On the other hand, for
a constant $\rm{^{56}Ni}$ mass a smaller recombination temperature
implies an increase in the ejecta velocity and mass, and therefore
in the explosion energy.

The later phase is not powered by $\rm{^{56}Ni}$ alone. The main
source of energy is in fact the transformation of the kinetic energy
of the ejecta into thermal energy and radiation inside the dense
clumps, which form a photosphere. The duration and shape of the
luminosity peak depends on the radius and mass of the clumps and on
their expansion velocity.  The parameters listed in Table 6 are the
clump radius, mass and velocity, the amount of $\rm{^{56}Ni}$ in the
clumps, the recombination temperature, the energy released by the
ejecta-clumps interaction and the diffusion time. The recombination
temperature adopted is $\rm{T=6500\pm1000}$K, as measured from the
37-day spectrum. For both models reported (\textit{c1} and
\textit{c2}) the energy deposited by the ejecta in the CSM is about
a factor $\sim$2-30 smaller than the SN explosion energy. This value
may result naturally, considering that the clumps cover a solid
angle not larger than 2$\pi$ as seen from the centre of the star. In
the two models the radius of the clump is significantly different.
For an ejecta velocity of 8000 $\rm{km\,s^{-1}}$, and assuming that
the ejecta-clump impact occurs at $\sim$30-40 days, the distance of
the clumps is $\rm{\sim10^{15}\,cm\,s^{-1}}$. Adopting a
characteristic sound speed of $\rm{\sim10^{8}cm^{-1}}$, the  shock
wave produced by the impact takes $\sim 100$\,days to cross the
clump  in model \emph{c1} and $\sim 10$\,days in model \emph{c2}. On
these grounds, model \emph{c2} seems to be favored, as the optical
display of the shocked clumps is fully developed by $\sim 40$ days
after explosion. The values obtained for the clump distance and mass
are roughly consistent with those derived by \citet{smcray07}.

Our simple model gives a satisfactory fit for both the explosion and
collision phase (Figure 8). We did not attempt to fit the light
curve in the transition phase. The parameters that characterize the
explosion of SN\,2006gy are actually not particularly remarkable. An
extra-ordinary amount of $\rm{^{56}Ni}$ in the ejecta is not
necessary to fit the light curve. The estimated amount of
$\rm{^{56}Ni}$ is 2 to 6 times larger than that derived for other
well studied, bright core-collapse events \citep{mt00,maz06}. It
should be noted that an high amount of $\rm{^{56}Ni}$ is even not to
relate to the huge brightness of the recently discovered SN\,2008es,
the second most luminous SN known, according to \citet{gez08} and to
\citet{mill08}. SN\,2006gy was certainly a highly energetic event
compared to other normal CC-SNe, perhaps comparable to the class of
\textit{hypernovae} (e.g., SN\,2003dh, \citealt{mh03}; SN\,2003jd,
\citealt{sv08}; SN\,1998bw, \citealt{iwa98}). The combined mass of
the ejecta and of the clumps is $\rm{\sim20M_{\odot}}$, indicating
an originally very massive progenitor, likely much more massive than
$\rm{\sim30M_{\odot}}$ if account is taken of the likely large mass
loss in the pre-SN stage. Still, these values are significantly
smaller than those claimed in some of the previously proposed
scenarios ($\rm{>100M_{\odot}}$).

There is increasing evidence for the association of bright SNe\,IIn
and LBVs \citep{sal02,smow06,kot06,gy07,tr08}. The properties of
these events seem to require that their progenitor stars experienced
mass loss rates of the order of $\rm{\sim0.1M_{\odot} yr^{-1}}$,
which are only compatible with those estimated for stars like $\eta$
Carinae. Also in this case strong, LBV-like mass-loss phenomena are
required to produce massive clumps around the star. Given the radius
at explosion derived by the model, a star in a LBV or early
Wolf-Rayet phase might be good candidates for the progenitor of
SN\,2006gy.

\section{Summary and conclusions}

New observational data of SN\,2006gy allow us to derive constraints
on the physical processes underlying SN 2006gy. We confirm the
luminosity drop in the R band at days 362 and 394 first reported by
S08, and find a similar drop also for the B and I bands at similar
epochs. The absence of SN features in the spectrum at
$\rm{t\sim389}$\,days supports this. In all bands the light curves
exhibit a broad luminosity peak at day $\sim 70$, followed by a
steep decline and then  by a flattening at $\sim$170-237 days. At
very late phases the SN was detected in the K band.  This may deal
either with dust formation in the ejecta or with IR echoes events.
However, the uncertainties on the late-phases scenario do not hamper
the estimate of $\rm{^{56}Ni}$ mass ejected. Based on the bolometric
luminosity at day $\sim$180 un upper limit of $\rm{\sim3M_{\odot}}$
is derived. At this epoch interaction, rather than radioactive
decay, has been proved to be the dominant source.

During the first 3 months the behavior of SN\,2006gy can be
reproduced as the explosion of a compact progenitor star (with
explosion energy $\rm{\sim 4-9\cdot10^{51}erg}$, $\rm{R \sim6-8\cdot
10^{12}cm}$). The SN ejecta collide with some previously ejected
material ($\rm{\sim6-10M_{\odot}}$) distributed in highly opaque
clumps. The increasing size of the SN ejecta, the relatively large
amount of $\rm{^{56}Ni}$ ejected, the collision with the extended,
opaque clumps with a long diffusion time are the ''ingredients''
responsible for the slow increase of the light curve to maximum and
for the brightness and extension of the peak. The values derived for
the mass of the clumps and their radial distance are consistent with
those derived for the shocked shell by \citet{smcray07}.

The spectra of SN\,2006gy at $\sim$170-237 days are similar to those
of a number of bright, interaction-dominated SNe (Figure 6), with
which SN\,2006gy shares remarkable photometric similarities (Figure
4). This confirms that at this epoch interaction plays a dominant
role also in the case of SN\,2006gy. In the massive-clump scenario,
it is likely that interaction signatures start to dominate the
spectra when the clumps become transparent because they recombine.
The ejecta encounter regions of progressively lower density with
time. At the epoch of our last optical observation ($\sim$423 days),
interaction has probably already ceased. This is consistent with the
non-detection of H${\alpha}$ in the Keck spectrum of S08.

In conclusion, although SN\,2006gy was very luminous and energetic
($\sim3$ to 4.5 times more energetic than SN\,1987A), it does not
appear to be an extraordinary event. In fact, neither the explosion
of a supermassive progenitor, nor extremely high Ni-rich ejecta are
required to explain the observations.

Unfortunately, the nature of the progenitor star of SN\,2006gy still
remains obscure. Nevertheless, to account for such a violent and
energetic explosion as well as for the existence of an extremely
dense CSM around the exploding star it is natural to consider a very
massive progenitor, likely more massive than $\rm{30M_{\odot}}$,
which experienced strong mass loss episodes just before the
explosion.

LBV-like mass loss rates resulting in highly dense circumstellar
shells seem to be common near the bright end of interacting
supernovae (e.g., SN\,2006jc, \citealt{apas06}; SN\,1997eg,
\citealt{hoff07}; SN\,2005gj, \citealt{tr08} and S. Benetti et al.
2008, in preparation). A LBV-like outburst was also claimed by
\citet{sm08b} to explain the observational data of SN\,2006tf.
Considering the radius of the progenitor required to explain the
long luminosity rise to maximum, the progenitor of SN\,2006gy was
probably a LBV or an early Wolf-Rayet star.

\acknowledgments We thank Nathan Smith and Dick Mc Cray for the
precious suggestions and comments to this paper. MT, EC and SB are
supported by the Italian Ministry of Education via the PRIN 2006 n.
022731-002. This manuscript is based on observations collected at
Asiago observatory (Italy), at Telescopio Nazionale Galileo (La
Palma, Spain) and at Nordic Optical Telescope (La Palma, Spain). It
is also based [in part] on data collected at Kiso observatory
(University of Tokyo) and obtained from the SMOKA, which is operated
by the Astronomy Data Center, National Astronomical Observatory of
Japan. The paper also makes use of data obtained from the Isaac
Newton Group Archive which is maintained as part of the CASU
Astronomical Data Center at the Institute of Astronomy, Cambridge.
This research has made use of the NASA/IPAC Extragalactic Database,
(NED), which is operated by the Jet Propulsion Laboratory,
Californian Institute of Technology, under contract with the
National Aeronautics and Space Administration. We acknowledge the
use of HyperLeda database, supplied by the LEDA team at the Centre
de Recherche Astronomique de Lyon, Observatoire de Lyon.





\clearpage

\begin{flushleft}
\begin{deluxetable}{cccccccc}
\setlength\tabcolsep{1.2pt} \tabletypesize{\small}
\tablecaption{Journal of photometric and spectroscopic observations
of SN\,2006gy.}
 \tablewidth{0pt}
\tablehead{ \colhead{UT Date} & \colhead{Telescope} &
\colhead{Equipment} & \colhead{Bands} & \colhead{Grisms} &
\colhead{Spec. range [{\AA}]} & \colhead{Resolution [{\AA}]} &
\colhead{Pixelscale  [''/pix.]} } \startdata
06/09/25 & TNG\tablenotemark{b} & DOLORES & - &  LR-B, LR-R & 3200-9000 & 18, 17 & 0.25  \\
06/09/30 & NOT\tablenotemark{a} & ALFOSC & BVRI & 4 & 3400-8800 & 21 & 0.19 \\
06/10/29 & Ekar1.82m & AFOSC & BVRI & - & - & - & 0.46 \\
06/12/19 & Ekar1.82m & AFOSC & BVRI & 4 & 3500-7500& 24 & 0.46  \\
07/02/10 & NOT & ALFOSC & BVRI & 4, 5 & 3500-9800 & 21,`20 & 0.19 \\
07/03/10 & Ekar1.82m & AFOSC & BVRI & - & - & -& 0.46  \\
07/03/12 & Ekar1.82m& AFOSC & - & 2, 4 & 3400-7600 & 38, 24 & 0.46  \\
07/04/13 & Ekar1.82m& AFOSC & VRI & - &- & -& 0.46  \\
07/09/14 & Ekar1.82m& AFOSC &  R & 4 & 3400-7600 & 24 & 0.46  \\
07/10/05 & TNG\tablenotemark{b}  & NICS & JHK' & - & - & - & 0.25  \\
07/10/17 & TNG\tablenotemark{b} & DOLORES & BRI & - & -& - & 0.25  \\
08/01/12  & TNG\tablenotemark{b}  & NICS & K' & - & - & - & 0.25  \\
\enddata
\tablenotetext{a}{Nordic Optical Telescope}
\tablenotetext{b}{Telescopio Nazionale Galileo}
\end{deluxetable}
\end{flushleft}

\begin{deluxetable}{ccccccc}
\setlength\tabcolsep{1.2pt} \tabletypesize{\small}
\tablecaption{Main data on the archive images used for the
photometric template subtraction.}
 \tablewidth{0pt}
\tablehead{ \colhead{UT Date} & \colhead{Telescope} &
\colhead{Equpment} & \colhead{Bands} & \colhead{Exp.time} &
\colhead{Seeing [{''}]}  & \colhead{Pixelscale [''/pix.]} }
\startdata
91/12/01 & JKT\tablenotemark{c} & AGBX & B & 600 & 1 & 0.33\\
91/12/01 & JKT & AGBX & R &300 & 1.15 & 0.33 \\
96/01/13 & JKT & AGBX & I & 360 & 1.5 & 0.33 \\
03/02/12 & Schmidt Tel.&  & V & 300 & 3.7 & 1.46 \\
\enddata
\tablenotetext{c}{Jakobus Kapteyn Telescope}
\end{deluxetable}

\clearpage

\begin{deluxetable}{ccccccccccc}
\setlength\tabcolsep{1.2pt} \tabletypesize{\small}
\tablecaption{Optical photometry of SN\,2006gy.} \tablewidth{0pt}
\tablehead{ \colhead{UT Date} & \colhead{JD -2,400,000} &
\colhead{Phase [days]\tablenotemark{a}} & \colhead{B} &
\colhead{Berr} & \colhead{V} & \colhead{Verr} & \colhead{R} &
\colhead{Rerr} & \colhead{I}& \colhead{Ierr}} \startdata
06/09/18 & 53996.5 & 29.5 & - & -  &  -  & - &  15\tablenotemark{b} & - & - & -\\
06/09/30 & 54008.5 & 41.5 & 16.00 &  .08 & 15.19 & .22 & 14.51 & .10 & 14.48& .07 \\
06/10/29 & 54037.5 & 70.5 & 15.84 & .06 & 14.85 & .13 & 14.28 & .04 & 14.10 & .06 \\
06/12/19 & 54088.5 & 121.5 & 16.23& .06 &15.08 &.15 &14.99& .04& 14.37 &.06 \\
07/02/10 & 54142.4 & 174.5 & 17.87 &.08& 16.75& .22& 16.69 &.10& 15.86& .07\\
07/03/10 & 54169.5 & 204.5 & 18.07 & .06& 17.74 &.15 & 16.67 & .04& 16.19 &.06 \\
07/04/13 & 54203.5 & 236.5 & - & - &17.66&  .15 &16.93& .04& 16.32& .06\\
07/09/14 & 54356.5 & 389.5 & - & - & - & - & $>$20.30 & -& - & - \\
07/10/17 & 54390.6 & 423.5 & $>$21 & - & - & - & $>$21.55 & - & $>$19.75 & - \\
\enddata
\tablenotetext{a}{With respect to
JD=2453967.0}\tablenotetext{b}{From ATEL 644 (2006)}
\end{deluxetable}

\begin{deluxetable}{c c c c c c c c c}
\setlength\tabcolsep{1.2pt} \tabletypesize{\small}
\tablecaption{Near--infrared photometry of SN
2006gy.}\tablewidth{0pt} \tablehead{ \colhead{UT Date} & \colhead{JD
-2,400,000} & \colhead{Phase [days]\tablenotemark{a}} & \colhead{J}
& \colhead{Jerr} & \colhead{H} & \colhead{Herr} & \colhead{K} &
\colhead{Kerr} } \startdata
07/10/05 & 54378.5 & 411 & $>17$ &  - & $>16.5$ & - & 16.00 & .50  \\
08/01/12 & 54477.5 & 510 & - & - & - & - & 16.3 & .50 \\
\enddata
\tablenotetext{a}{With respect to JD=2453967.0}
\end{deluxetable}

\clearpage

\begin{deluxetable}{c c c c c c c}
\setlength\tabcolsep{1.2pt} \tabletypesize{\small}
\tablecaption{Model output parameters of the semi-analytical code
for the first evolutionary phase.}\tablewidth{0pt} \tablehead{
\colhead{model} & \colhead{$\rm{R_{star}[\cdot10^{12}cm]}$} &
\colhead{$\rm{M_{ej}[M_{\odot}]}$} & \colhead{$\rm{V_{ej}[km/s]}$} &
\colhead{$\rm{M_{Ni}[M_{\odot}]}$} & \colhead{$\rm{T_{rec}[K]}$} &
\colhead{$\rm{E_{expl}[10^{51}erg]}$} } \startdata
(\textit{e1}) & 8.4  &  5.3 & 7700  & 0.75  &  7000  &  3.8 \\
(\textit{e2}) & 5.9& 8.3  & 8900 &  1.0  &  6500 &   7.9 \\
(\textit{e3})  & 8.4 & 6.9 & 7700 &  1.0 & 7000 &  4.9 \\
(\textit{e4})  & 8.4 & 14.4 & 7300 &  2.0 & 7000 &  9.2 \\
\enddata
\end{deluxetable}

\begin{deluxetable}{c c c c c c c c}
\setlength\tabcolsep{1.2pt} \tabletypesize{\small}
\tablecaption{Model output parameters of the semi-analytical code
for the second evolutionary phase.}\tablewidth{0pt} \tablehead{
\colhead{model} & \colhead{$\rm{R_{cl}[\cdot10^{12}cm]}$} &
\colhead{$\rm{M_{cl}[M_{\odot}]}$} & \colhead{$\rm{V_{cl}[km/s]}$} &
\colhead{$\rm{M_{Ni}[M_{\odot}]}$} & \colhead{$\rm{T_{rec}[K]}$} &
\colhead{$\rm{E_{imp}[10^{51}erg]}$}  & \colhead{diff. time
[$\rm{days}$]} } \startdata
(\textit{c1}) & 1339.9 &  6.5  &  1900  &   0.1   &  6500 & 0.3  & 100 \\
(\textit{c2}) & 290.6  &  10.0  &  3600  &   0.1   &  6500 &  1.6   & 10 \\
\enddata
\end{deluxetable}

\clearpage

\begin{figure}
\epsscale{.80} \plotone{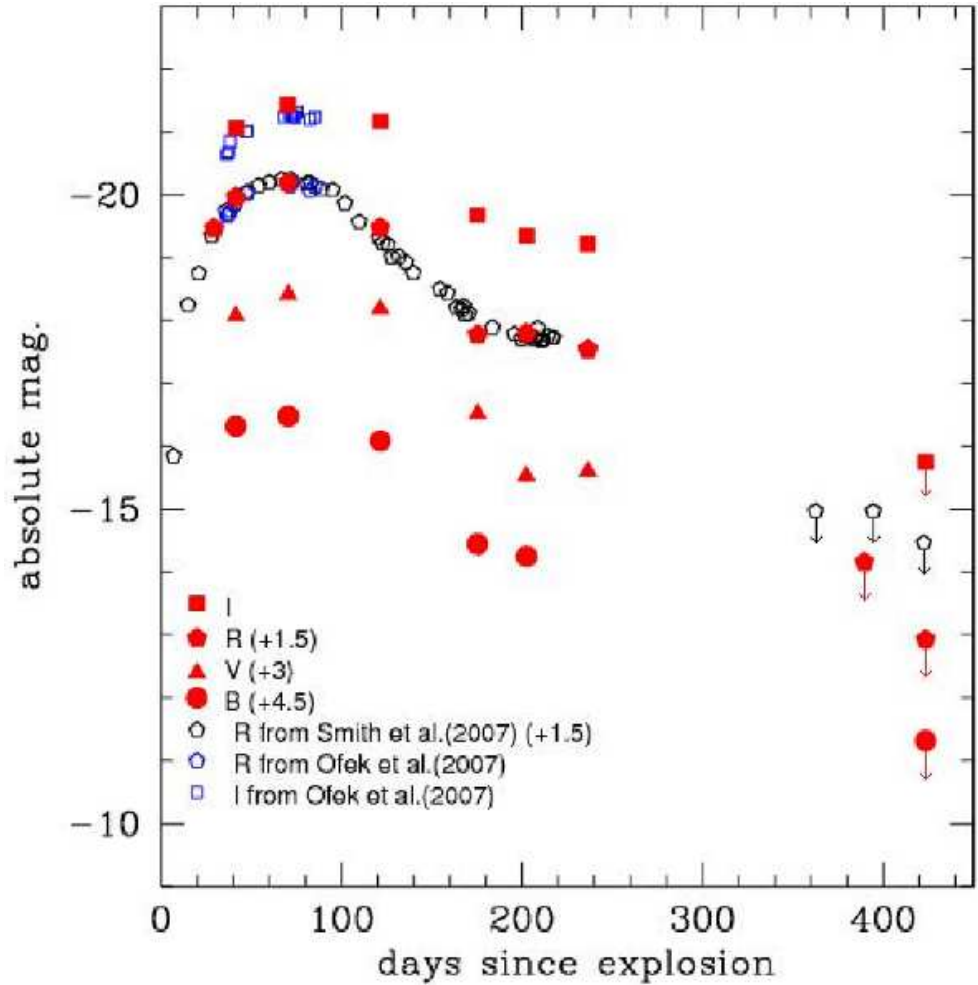}
 \caption{BVRI absolute light curves
of SN\,2006gy, obtained with the distance and extinction reported in
the text. Late phase ($>$300 days) detection limits are marked with
an arrow. R data from \citet{sm07} and \citet{sm08}, as well as R
and I data from \citet{of07} are also reported.\label{fig1}}
\end{figure}

\clearpage

\begin{figure}
\epsscale{.90} \plotone{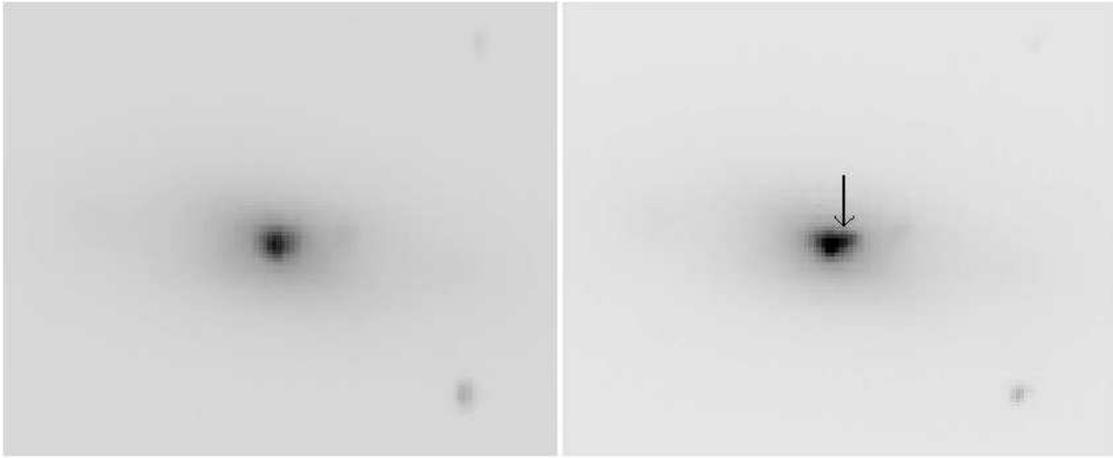} \caption{Images aquired with NICS at
TNG
 with filter J (left panel) and K' (right panel) on October 5th, 2007 (JD 2454378.5). SN 2006gy
 is still clearly visible near the host galaxy nucleus in the K' band image,
 whereas there is no detectable source at that position in the J band frame. \label{fig1}}

\end{figure}

\clearpage

\begin{figure}
\epsscale{.90} \plotone{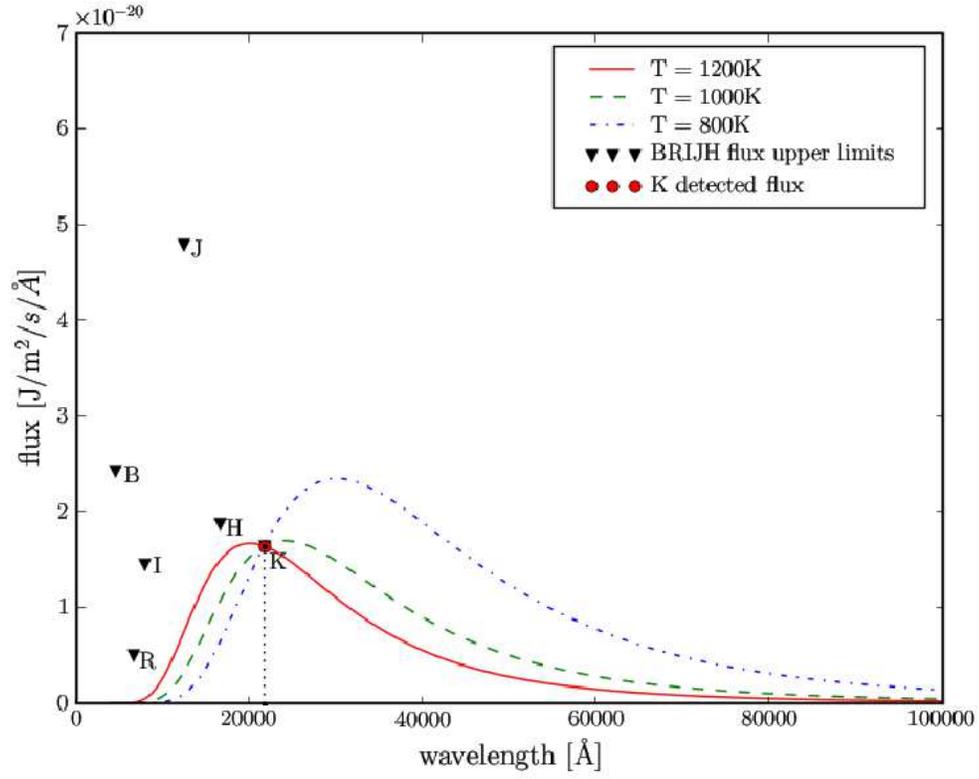} \caption{Comparison between the
optical (day 423) and near--infrared (day 411) flux measured for
SN\,2006gy. We also show the expected emission from dust at
different temperatures, normalized to the K band
magnitude.\label{fig1}}
\end{figure}

\clearpage

\begin{figure}
\epsscale{.95} \plotone{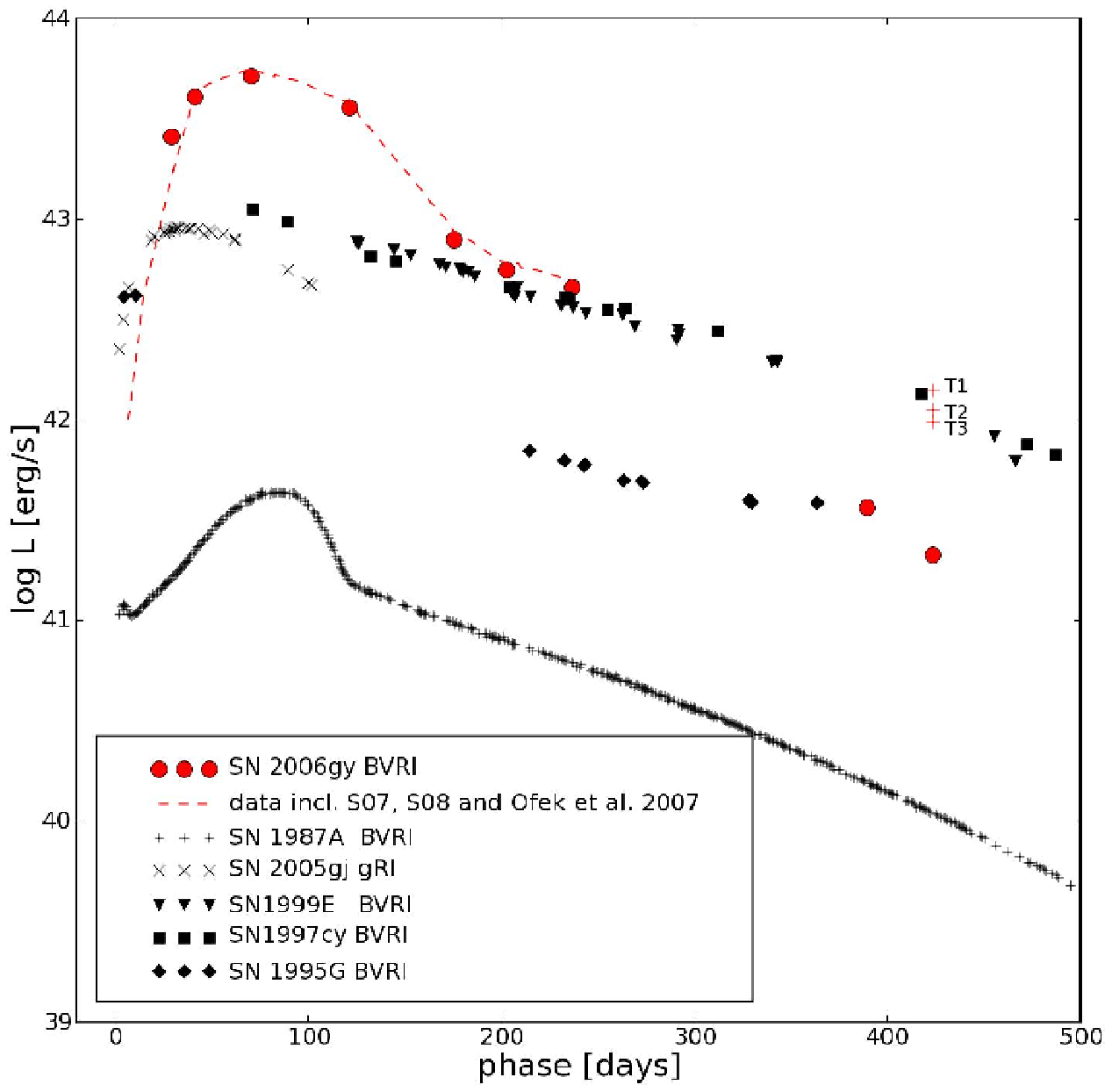} \caption{Pseudo-bolometric light
curve of SN\,2006gy compared to those of type IIP SN\,1987A
\citep{white87}, type IIn SN\,2005gj \citep{pr07}, SN\,1999E
\citep{rig03}, SN\,1997cy \citep{mt00,ger00} and SN\,1995G
(Pastorello et al. 2002), all integrated in the same wavelength
range. Red crosses at late times include the near-IR contribution
due to a possible cold dusty region in SN\,2006gy ejecta, based on
the K-band detection and on three possible dust temperatures
(T1=800K, T2=1000K and T3=1200K, see $\S3.1$). For SN\,1997cy and
SN\,1999E the epochs of the associated GRB explosions (GRB 970514
and GRB 980919) are adopted as phase reference epochs. \label{fig1}}
\end{figure}

\clearpage

\begin{figure}
\epsscale{.90} \plotone{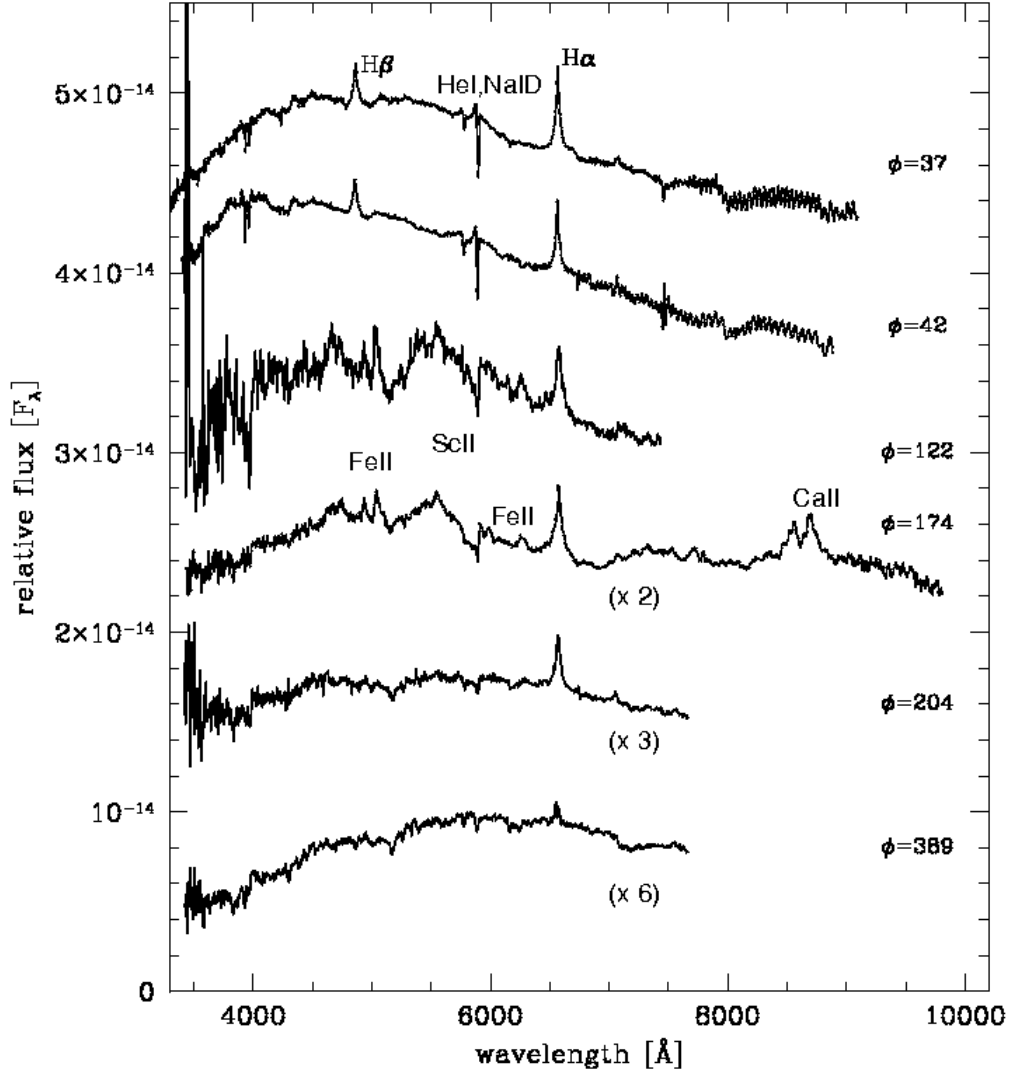}
 \caption{Spectroscopic evolution of
SN\,2006gy from 37 days to 389 days since explosion in the host
galaxy rest-frame, corrected for extinction assuming E(B$-$V)=0.56.
The spectra at phase 174, 204 and 389 were multiplied by a factor 2,
3 and 6 respectively.\label{fig1}}
\end{figure}

\clearpage

\begin{figure}[h]
\begin{center}
\includegraphics[height=7cm, keepaspectratio]{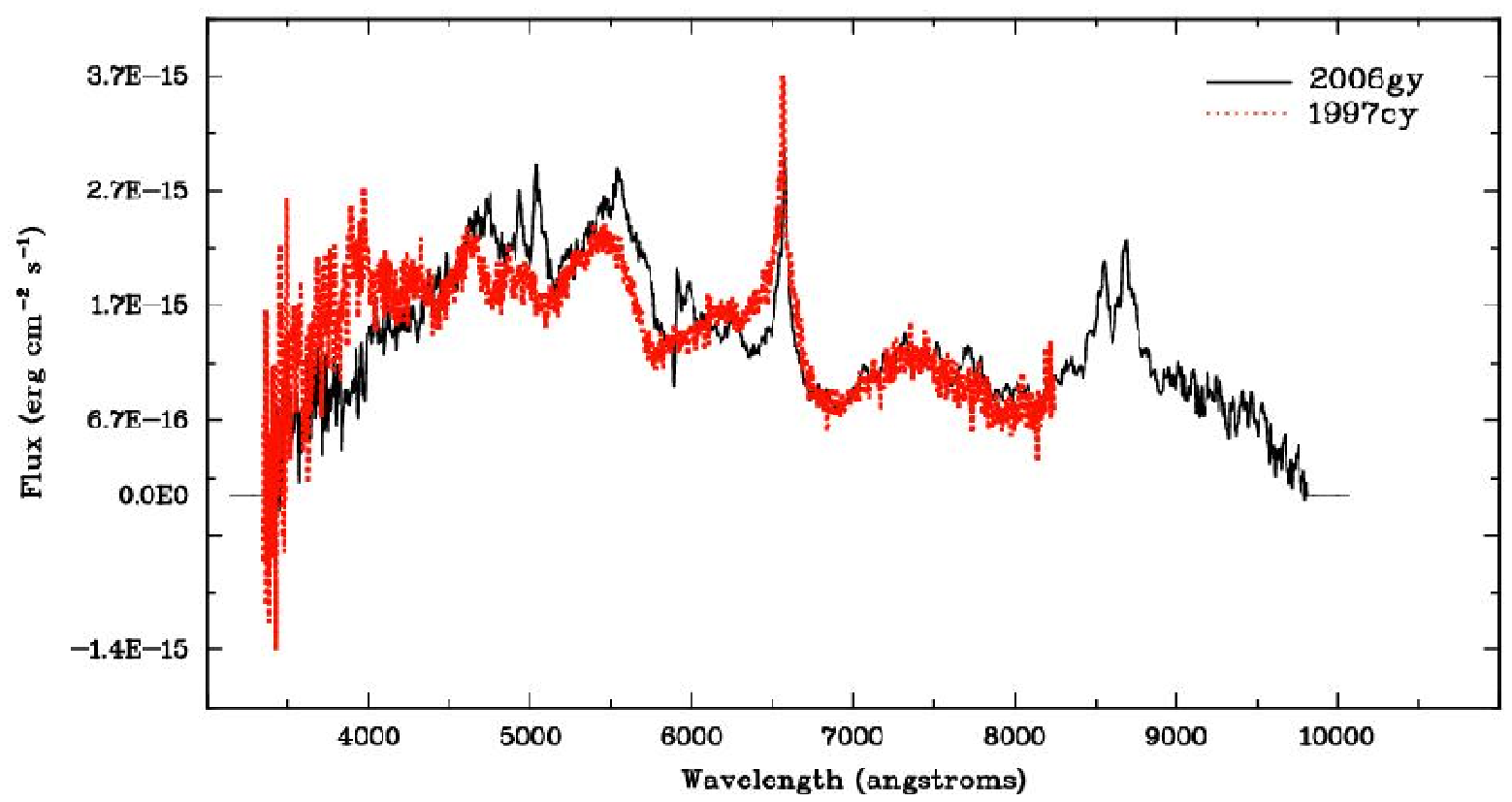}
\qquad\qquad
\includegraphics[height=7cm, keepaspectratio]{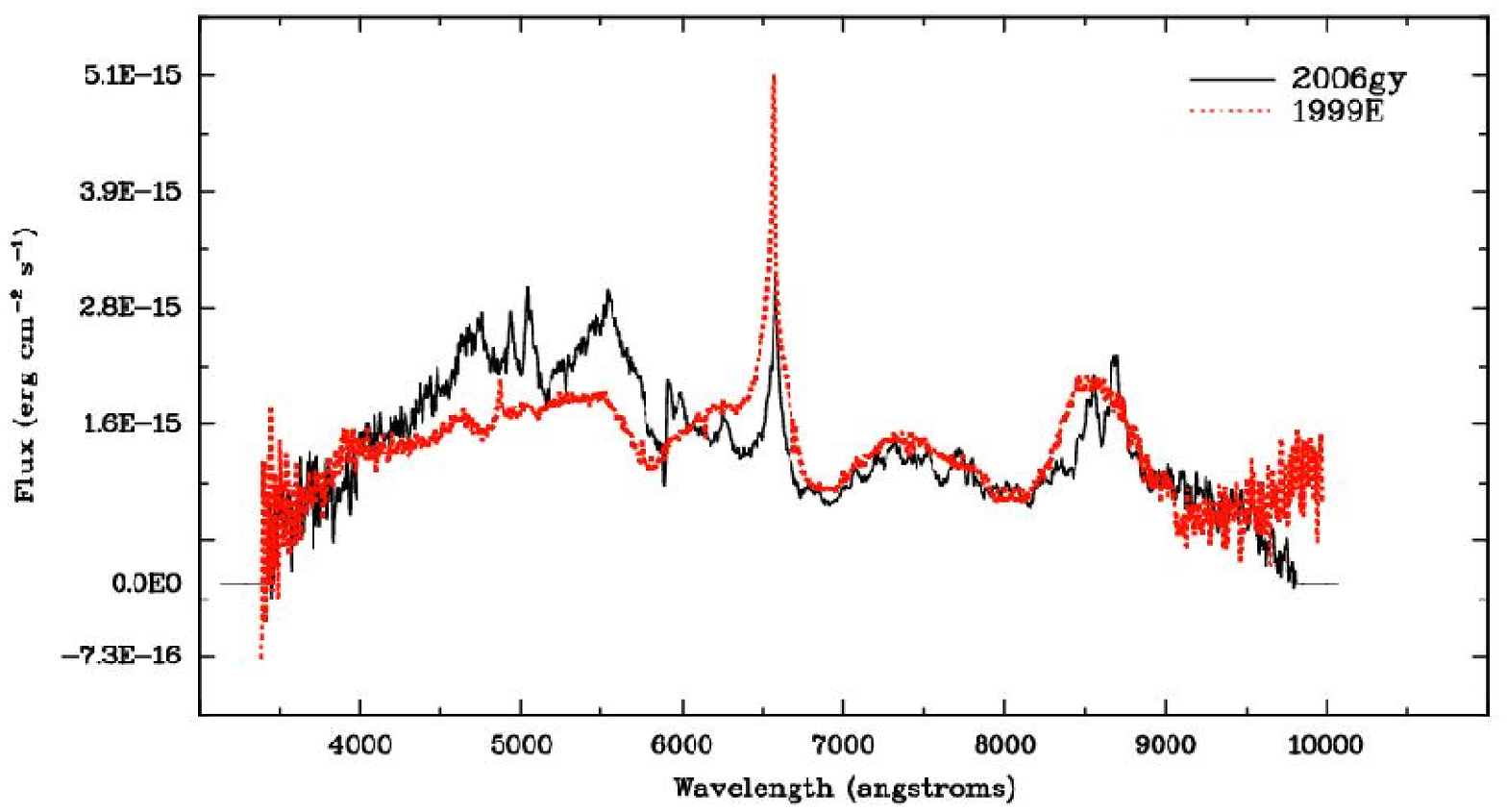}
\caption{GELATO comparison between the spectrum of SN\,2006gy at
phase $\sim$174 days, SN\,1997cy (top, \citealt{mt00}) and SN\,1999E
(bottom, \citealt{rig03}) at similar phases. Although the comparison
SNe have broader lines, (e.g., $\rm{FWHM_{H{\alpha}}}$=12800
$\rm{km\, s^{-1}}$ in SN\,1997cy according to \citealt{mt00}), the
objects show an overall remarkable similarity. \label{fig1}}
\end{center}
\end{figure}

\clearpage

\begin{figure}
\epsscale{.90} \plotone{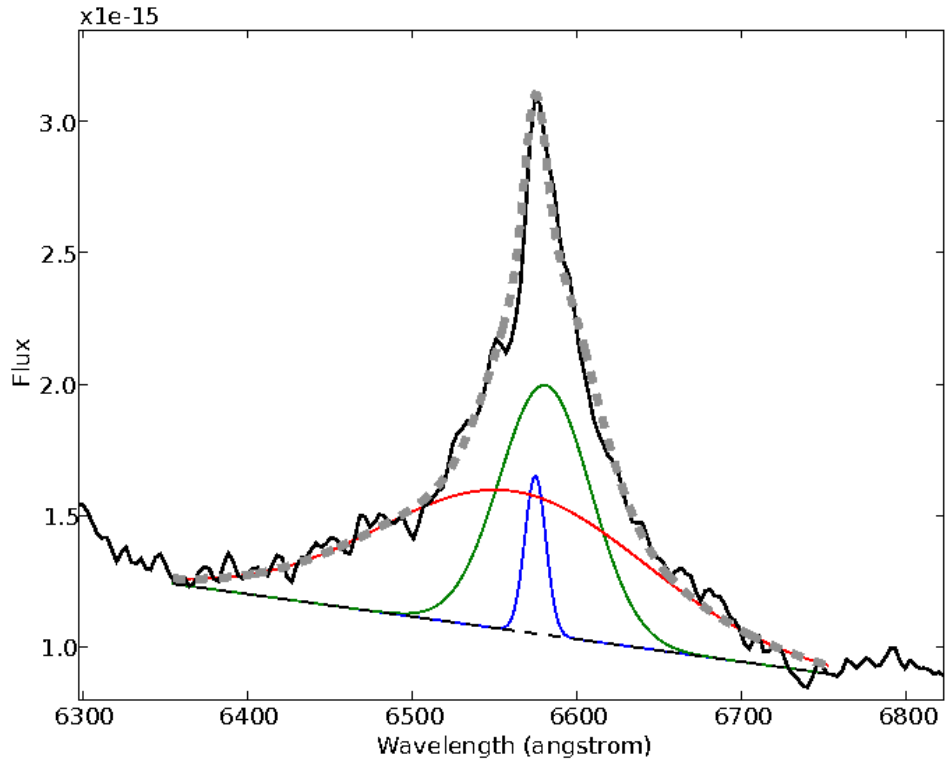} \caption{Detail of the $\rm{H\alpha}$
profle in the spectrum obtained at NOT on February 10th, 2007. The
line is decomposed into three gaussian profiles, having FWHM = 685
(unresolved), 3200 and 9000 $\rm{km\, s^{-1}}$.\label{fig1}}
\end{figure}

\clearpage

\begin{figure}
\epsscale{.90} \plotone{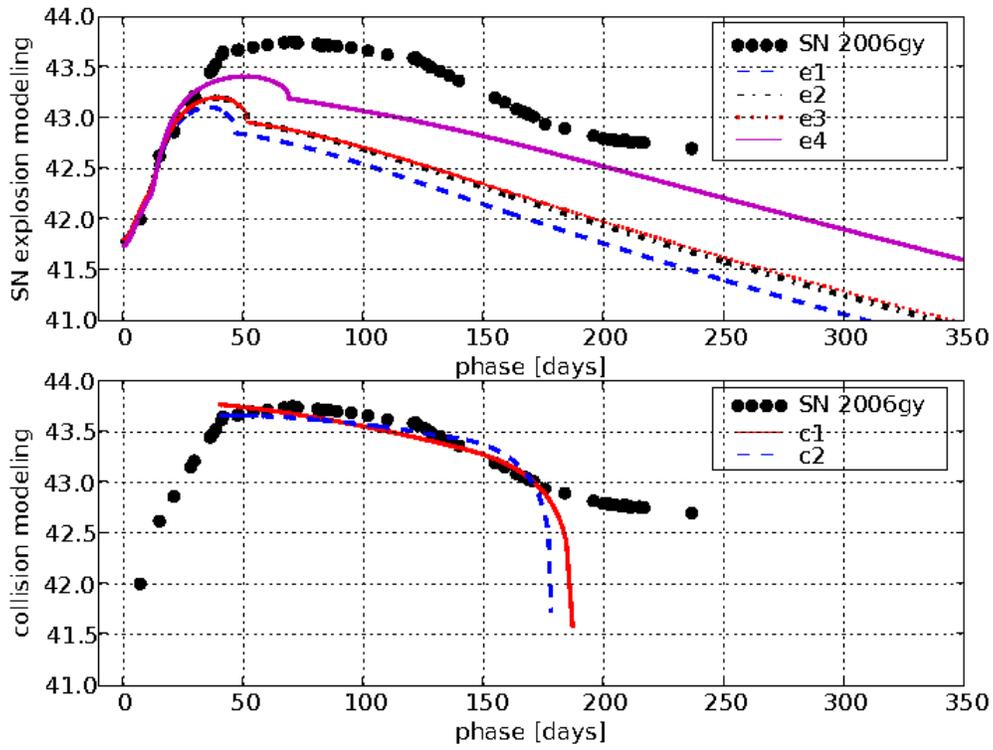} \caption{Best fits of the light curve
of SN\,2006gy obtained with the semi-analytical model \citep{lz03},
showed separately for the rising branch and maximum/post maximum
phase. The code in the legenda refers to the models summarized in
Table 5 and 6.\label{fig1}}
\end{figure}

\end{document}